\documentclass[conference,a4paper]{IEEEtran}

\usepackage[english]{babel}

\usepackage{amsmath,amssymb,amsfonts,mathrsfs}

\usepackage[amsmath,thmmarks]{ntheorem}

\usepackage{graphicx}

\newtheorem{protocol}{Protocol}
\newtheorem{theorem}{Theorem}
\newtheorem{lemma}{Lemma}
\newtheorem{corollary}{Corollary}

\begin{document}

\sloppy

\title{Distillation of Multi-Party Non-Locality \\With and Without Partial
Communication}

\author{
  \IEEEauthorblockN{Helen Ebbe}
  \IEEEauthorblockA{Department of Mathematics\\
    ETH Zentrum\\
    8092 Zurich, Switzerland\\
    Email: heebbe@ethz.ch} 
  \and
  \IEEEauthorblockN{Stefan Wolf}
  \IEEEauthorblockA{Faculty of Informatics\\
    University of Lugano (USI)\\
     6900 Lugano, Switzerland\\
    Email: wolfs@usi.ch}
}



\maketitle

\begin{abstract}
Non-local correlations are one of the most fascinating consequences
of quantum physics from the point of view of information: Such
correlations,
although not allowing for signaling, are unexplainable by pre-shared
information. The correlations have applications in cryptography,
communication complexity, and sit at the very heart of many attempts
of understanding quantum theory | and its limits | better in terms
of classical information. In these contexts, the question is crucial
whether
such correlations can be {\em distilled}, i.e., whether weak
correlations can be used for generating (a smaller amount of)
stronger.
Whereas the question has been studied quite extensively for bipartite
correlations (yielding both pessimistic and optimistic results), only
little is known in the {\em multi-partite\/} case.

We show that a natural generalization of the well-known
Popsecu-Rohrlich box can be distilled, by an adaptive protocol,
to the algebraic maximum. We use this result further to show that
a much bigger class of correlations, including \emph{all} purely three-partite correlations, can be distilled from
arbitrarily weak to maximal strength with partial communication,
i.e.,
using only a subset of the channels required for the creation of the
same correlation from scratch. In other words, we show that arbitrarily
weak non-local correlations can have a ``communication value''
in the context of the generation of maximal non-locality.
\end{abstract}

\section{Introduction}
One of the most mysterious, challenging, but also useful consequences
of quantum theory is the possibility of non-local correlations: The
joint
behavior under (different possible) measurements of a quantum system
is such that it cannot be explained by pre-shared (classical)
information
determining all the outcomes locally. This result by Bell~\cite{bell}
can be seen as a late reply to the claim, in 1935, of Einstein,
Podoslky, and Rosen~\cite{EPR} that quantum theory was incomplete and
must be augmented by {\em hidden variables}, i.e., classical
information predicting all measurements' outcomes.\footnote{Bell's
 result only persists under the assumption that measurement bases are
chosen freely; but at the same time, none of the deterministic
interpretations of quantum physics satisfies with an explanation of
the
correlations' origin.}

It has been a prominent open problem ``why'' nature does
display non-local behavior, yet no maximal one, i.e., the behavior of a
perfect PR box~\cite{pr} cannot be realized~\cite{cir}.
A number of attempts have been made to single out quantum correlations
as compared to general non-signaling systems: Are quantum correlations
the ones that do not collapse {\em communication complexity\/}~\cite{cc}, that
are of no help for {\em non-local computation\/}~\cite{nlc}, or that respect
{\em information causality}, a principle generalizing the non-signaling
principle to the case of limited communication~\cite{ic}?
Furthermore, it has turned out that non-local correlations have
important applications for information processing, e.g.,
device-independent cryptography or communication complexity. In all
the mentioned contexts, a question of paramount importance is the one
of
{\em distillation of non-locality\/}: Given weak correlations, is it
possible to generate stronger by some local wirings?
For instance, distillation can potentially lead to higher
confidentiality levels or to a collapse of communication
complexity by apparently weak correlations.

In the two-party scenario, the possibility of distillation has already
been extensively studied and, notably, led to complementary results
adding up to a pretty complex picture: Whereas {\em isotropic
 CHSH-type}~\cite{CHSH}
correlations seem undistillable~\cite{dukwol}, the same fails to be
true in general~\cite{FWW09}, \cite{BS09}, \cite{HoyerRashid}. In fact,
certain
arbitrarily weak CHSH correlations can even be distilled up to
arbitrarily close
to perfect PR boxes by adaptive protocols.

In the case of three or more parties, much less is known. It was shown
that the straight-forward generalization of the (non-adaptive) XOR
protocol~\cite{FWW09}
to more parties fails to distill extremal boxes of the non-signalling polytope to
almost-perfect~\cite{hsuwu}.

The contribution of the present work is two-fold: First, we show that
the natural generalization of PR boxes to $n$ parties has the property
that non-isotropic faulty versions of it can be distilled to
close-to-perfect
by a multi-party variant of the BS protocol (Section III). Second, this result is
used to show distillability for a much larger class of correlations,
where the distillation is supported by partial communication, i.e., a
subset of the parties is allowed to communicate, where this communication
{\em alone\/} is insufficient for generating the target correlation
(Section IV).
This result can alternatively be interpreted as arbitrarily weak
non-local correlations having a ``communication value'' in the context
of the generation of almost-perfect systems.
In Section~V, the general results and procedures are illustrated with
a representative example.

\section{Definitions}
Here we define certain classes and specific types of $n$-partite boxes which we will use in our distillation protocols. They are generalizations of important bipartite boxes in \cite{FWW09,BS09,BP05}.

The most general type that we define is a \emph{full-correlation box}. Intuitively speaking it has a correlation only w.r.t. the \emph{full} set of players.
A \emph{full-correlation box} is an $n$-partite box which takes $n$ inputs and produces $n$ outputs. We denote the $n$-tuple of inputs as $\vec{x} = (x_1,x_2,...,x_n)$, where $x_i \in \lbrace 0, 1\rbrace$. The $n$-tuple of outputs is $\vec{a} = (a_1,a_2,...,a_n)$, where $a_i \in \lbrace 0, 1\rbrace$ for all $i$. The \emph{full-correlation box} is characterized by the following conditional distribution:
\begin{equation}
P(\vec{a}\vert \vec{x}) = \begin{cases} \frac{1}{2^{n-1}}&\text{$\sum\limits_i a_i \equiv f(\vec{x})$ (mod 2)}\\0&\text{otherwise,}\end{cases}
\end{equation}
where $f(\vec{x})$ is a Boolean function of the inputs. Two special cases of this type of box are the \emph{$n$-partite Popescu-Rohrlich box} and the \emph{even parity box for n parties}.
An \emph{$n$-partite Popescu-Rohrlich box} (or short \emph{n-PR box}) takes $n$ inputs $\vec{x} =  (x_1,x_2,...,x_n)$ and produces $n$ outputs $\vec{a} = (a_1,a_2,...,a_n)$ according to the conditional distribution
\begin{equation}
P^{PR}_n(\vec{a}\vert \vec{x}) = \begin{cases} \frac{1}{2^{n-1}}&\bigoplus\limits_{i}a_i = \prod\limits_{i} x_i\\0&\text{otherwise.}\end{cases}
\end{equation}
An \emph{even-parity box for n parties} takes $n$ inputs $\vec{x} = (x_1,x_2,...,x_n)$ and produces $n$ outputs $\vec{a} = (a_1,a_2,...,a_n)$ according to the conditional distribution
\begin{equation}
P_n^c(\vec{a}\vert \vec{x}) = \begin{cases} \frac{1}{2^{n-1}}&\bigoplus\limits_{i}a_i = 0\\0&\text{otherwise.}\end{cases}
\end{equation}
Note that this latter box is \emph{local}.
A convex combination of the last two boxes is called a \emph{correlated non-local box for n parties}.
The \emph{family of correlated non-local boxes for n parties} is defined as follows:
\begin{equation}
P^{PR}_{n,\varepsilon} = \varepsilon P^{PR}_n + (1-\varepsilon) P^c_n
\end{equation}
where $0\leq\varepsilon\leq1$.

\section{Generalization of the Brunner-Skrzypczyk Protocol}

Brunner and Skrzypczyk presented in \cite{BS09} a protocol for two parties that distills non-locality and in the asymtotic limit: All correlated non-local boxes are distilled to the maximally non-local PR box. This result can be generalized to all $n$-partite PR boxes Protocol~\ref{prot}.

\begin{figure}[!htb]
\begin{center}
\includegraphics[width=0.6\linewidth]{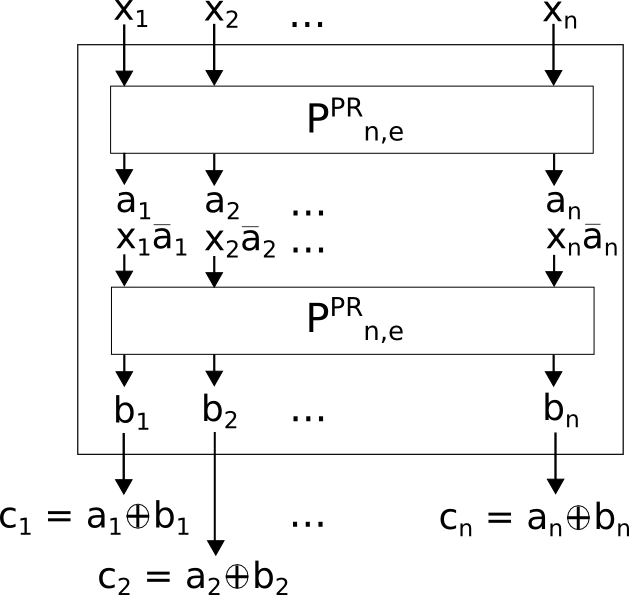} 
\end{center}
\caption[$n$-PR Box destillation]{Generalized BS protocol for $n$-PR boxes}
\label{fig:dbsprot}
\end{figure}

\begin{protocol}[Generalized BS Protocol for $n$-PR Boxes]
The protocol works as follows (see also Fig.~\ref{fig:dbsprot}). All n parties share two boxes, where we denote by $x_i$ the value that the $i$th party inputs to the first box and by $y_i$ the value that the $i$th party inputs to the second box. The output bit of the first box for the $i$th party is then $a_i$, and the output bit of the second box is $b_i$. The n parties proceed as follows: $y_i =  x_i\bar{a}_i$ and they output, finally, $c_i = a_i \oplus b_i$.
\label{prot}
\end{protocol}

With this protocol we are also able to distill a large class of boxes arbitrarily closely to the $n$-PR box.

\begin{theorem}
\label{thm:bsprot}
The generalized BS protocol takes two copies of an arbitrary box $P^{PR}_{n,\varepsilon}$ with $0<\varepsilon<1$ to an $n$-partite correlated non-local box $P^{PR}_{n,\varepsilon'}$ with $\varepsilon'>\varepsilon$, i.e., is distilling non-locality. In the asymptotic case of many copies, any $P^{PR}_{n,\varepsilon}$ with $0<\varepsilon$ is distilled arbitrarily closely to the n-PR box.
\end{theorem}

Since the Protocol~\ref{prot} and Theorem~\ref{thm:bsprot} are generalizations of~\cite{BS09}, the proof works almost in the same manner.

\begin{IEEEproof}
We start with the initial two-box state of the protocol which is given by
\begin{eqnarray}
P^{PR}_{n,\varepsilon}P^{PR}_{n,\varepsilon} & = &\varepsilon^2 P^{PR}_{n}P^{PR}_{n} +(1-\varepsilon)^2 P^{c}_{n}P^{c}_{n}\nonumber \\& & + \varepsilon\left( 1-\varepsilon\right)\left( P^{PR}_{n}P^{c}_{n} + P^{c}_{n}P^{PR}_{n} \right)\ .
\end{eqnarray}

We apply the above distillation protocol and get the final box. As in \cite{BS09}, we use the notation $P_iP'_i \longrightarrow P_f$, which means that the protocol takes two initial boxes, $P_i$ and $P_i'$, to one copy of the final box $P_f$.

So we get the following relations: $P^{PR}_{n}P^{PR}_{n}\longrightarrow P^{PR}_{n}$, $P^{PR}_{n}P^{c}_{n}\longrightarrow P^{PR}_{n}$, $P^{c}_{n}P^{PR}_{n}\longrightarrow 2^{1-n}P^{PR}_{n} + \left( 1-2^{1-n}\right) P^{c}_{n}$, and $P^{c}_{n}P^{c}_{n}\longrightarrow P^{c}_{n}$.

After the application of the distillation protocol we get the final box, which is given by
\begin{eqnarray}
P^{PR}_{n,\varepsilon'} & = & \frac{\varepsilon}{2^{n-1}}\left( 2^{n-1} + 1 - \varepsilon\right) P^{PR}_{n}\nonumber \\ & & + \left( 1 - \frac{\varepsilon}{2^{n-1}}\left( 2^{n-1} + 1 - \varepsilon\right)\right)  P^{c}_{n}\ .
\end{eqnarray}
Hence, $\varepsilon' = \frac{\varepsilon}{2^{n-1}}\left( 2^{n-1} + 1 - \varepsilon\right) $. We are now able to determine what kind of  boxes can be distilled by this protocol. If the protocol distills the box $P^{PR}_{n,\varepsilon}$ to $P^{PR}_{n,\varepsilon'}$ then $\varepsilon$ has to fulfill  $\varepsilon'>\varepsilon$. We observe that all $0<\varepsilon<1$ fulfill this condition and, therefore, the protocol distills any box of the family of correlated non-local  boxes.

We show that in the asymptotic regime of many copies, any $P^{PR}_{n,\varepsilon}$ with $0<\varepsilon<1$ is distilled arbitrarily closely to the $n$-PR box. We are starting with $2^m$ copies of the box $P^{PR}_{n,\varepsilon}$ and get, finally, the box $P^{PR}_{n,\varepsilon_m}$, where $\varepsilon_m$ is the $m$th iteration of the map 
\begin{equation}T_n(\varepsilon) = \frac{\varepsilon}{2^{n-1}}\left( 2^{n-1} + 1 - \varepsilon\right) .
\end{equation}
The fixed points of this map are $\varepsilon = 0$ and $\varepsilon = 1$. To analyze the stability of these two fixed points we calculate the eigenvalues of the Jacobian (since the map is one-dimensional, the Jacobian is a real value and not a matrix). For the box $P^{c}_{n}$ ($\varepsilon = 0$), we find $\frac{dT}{d\varepsilon}\vert_{\varepsilon=0} = 1 + \frac{1}{2^{n-1}} > 1$, so this box is repulsive. For the other box $P^{PR}_{n}$ we find $\frac{dT}{d\varepsilon}\vert_{\varepsilon = 1} = 1 + \frac{1}{2^{n-1}} - \frac{1}{2^{n-2}} < 1$, so this box is attractive.
\end{IEEEproof}

\section{Application: Distillation with Partial Communication}
The generalized BS protocol can be used for distillation protocols for full-correlation boxes, where the use of communication is allowed to some of the parties. That means we are looking for distillation protocols based on the generalized BS protocol, but that are also allowing one-way communication channels between some of the parties, that can be used as often as required. We show that we are able to distill a general class of full-correlation boxes arbitrarily closely to the maximum with such a protocol.

\begin{lemma}
\label{lem:boolf}
If f is a Boolean function of the input elements $x_1,x_2,...,x_n$, then it can be written as
\begin{equation}
f(x_1,...,x_n) = \bigoplus\limits_{I \in \mathcal{I}} \left( a_I\cdot \bigwedge\limits_{i \in I}x_i \right) ,
\end{equation}
where $\mathcal{I} = \mathcal{P}\left( \lbrace 1,2,...,n \rbrace\right) $ and $a_I \in \lbrace 0, 1\rbrace$ for all $I \in \mathcal{I}$.
\end{lemma}

\begin{IEEEproof}
The constant, the AND, and the XOR allow for implementing the universal Boolean functions AND and NOT.
\end{IEEEproof}

\begin{figure}[!htb]
\begin{center}
\includegraphics[width=0.9\linewidth]{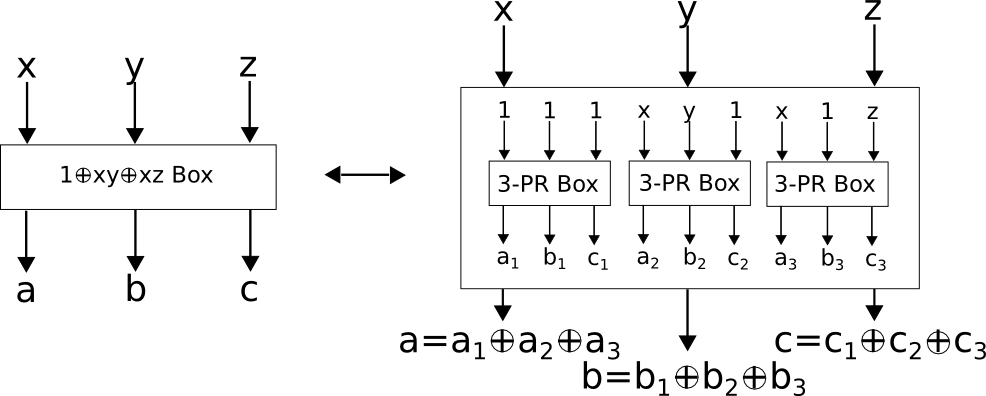} 
\end{center}
\caption[Construction of the $1\oplus xy \oplus xz$ box]{Construction of $1\oplus xy \oplus xz$ box}
\label{fig:equivalenz}
\end{figure}

Hence, it is obvious that the full-correlation box associated to the Boolean function $f$ can be constructed by $\sum\limits_{I \in \mathcal{I}}a_I$ $n$-PR boxes. Indeed, for every $a_I = 1$, an $n$-PR box is needed, where the $i$th party inputs $x_i$ if $i \in I$, and otherwise he inputs~$1$. Then, the box will output $b_i^I$. In the end, every party outputs $c_i = \bigoplus\limits_{I \in \mathcal{I},\ a_I =1}b_i^I$. For an example, see Fig. \ref{fig:equivalenz}. Note that the $n$-PR boxes belonging to $a_I$ where $\vert I \vert \leq 1$ are local and can be simulated by local operations and shared randomness.

We already know that all $n$-partite full-correlation boxes can be simulated by $n$-partite PR boxes. We define the set of all $n$-PR boxes that are needed to simulate the full-correlation box: Let

\begin{equation}
\mathcal{J} := \lbrace I \in \mathcal{I}\, \vert\, a_I = 1 \text{ and } \vert I\vert \geq 2\rbrace .
\end{equation}

This set can be partitioned into disjoint subsets $\lbrace J_1, J_2, ..., J_{n_\mathcal{J}}\rbrace$ such that all $A \in J_i$ and $B\in J_j$ fulfill $A\cap B = \emptyset$ for all $i \neq j$. We define the maximal number of such subsets as $n_{\mathcal{J}}$.
Later, we will see that it is important to know how many of the variables in a non-local box appear only in this non-local box, for that we define $m_I = \vert I \setminus \bigcup\limits_{J \in \mathcal{J}\setminus I} J \vert$ for all $I \in \mathcal{J}$.

Theorem \ref{thm:commcl} shows how many one-way communication channels are needed to simulate an $n$-partite full-correlation box.

\begin{theorem}[Number of one-way communication channels]
\label{thm:commcl}
Let f be the Boolean function associated to an $n$-partite full-correlation box, and let f be defined as in Lemma \ref{lem:boolf}. If $n_\mathcal{J}=1$, then the number $N^{scratch}_{comm}$ of one-way communication channels to simulate the full-correlation box from scratch is
	\begin{equation}
	N_{comm}^{scratch} =  \left|\bigcup\limits_{I\in \mathcal{J}} I \right| - 1 .
	\end{equation} 
\end{theorem}

\begin{IEEEproof}
We prove the statement by induction. We ignore the local part of the Boolean function $f$ (i.e. the terms of single variables) and start with the case when the function $f$ depends on two variables. The case $\vert\mathcal{J}\vert =2$ is equivalent to a PR-box. From \cite{PBS11}, we know that it can be simulated by one one-way communication channel. Now we assume that the claim is true for $\vert\mathcal{J}\vert \leq n$. Assume we have a function with  $\vert\mathcal{J}\vert =n+1$ that still fulfills the assumption of the theorem. We substitute $1$ for $x_i$, where $x_i$ is the input which is an element of a minimal number of elements of $\mathcal{J}$. This new function still fulfills the assumption of the theorem. We also know that $\vert\mathcal{J}\vert =n$ and, therefore, we need $n-1$ communication channels to simulate the associated box. We combine all these $n$ function values into one variable. The original function can be written with two variables. Therefore, we are back in the case $\vert\mathcal{J}\vert =2$. Together, we need $n$ one-way communication channels to simulate a function with $\vert\mathcal{J}\vert =n+1$.
\end{IEEEproof}

We construct an $n$-partite box where the outputs depend on the outputs of two full-correlation boxes for less than $n$ parties. These two boxes are defined by 
\begin{equation}
P_1(a_1 ...a_{k_2}\vert x_1 ... x_{k_2}) = \begin{cases}\frac{1}{2^{k_2-1}}& \bigoplus\limits_{i=1}^{k_2} a_i = g_1(x_1, ..., x_{k_2})\\0&\text{otherwise,}\end{cases}
\end{equation} 
where $g_1$ is a Boolean function which depends on all of its input variables and $k_2<n$. The second box is defined as
\begin{equation}
P_2(b_{k_1}...b_n\vert x_{k_1}...x_n) = \begin{cases}\frac{1}{2^{n-k_1}}& \bigoplus\limits_{i=k_1}^{n} b_i = \prod\limits_{i=k_1}^{k_3}x_i \\0&\text{otherwise,}\end{cases}
\end{equation} 
where $0<k_1<k_2<k_3\leq n$. These two boxes can be calculated in parallel.
Finally the constructed box outputs to party $i$
\begin{equation}
c_i = \begin{cases}a_i&  i \in\lbrace 1,2, ..., k_1-1\rbrace \\ a_i \oplus b_i & i \in\lbrace k_1,k_1+1, ..., k_2\rbrace \\b_i &  i \in\lbrace k_2+1, k_2+2, ..., n\rbrace . \end{cases}
\end{equation}

\begin{lemma}
\label{lem:constr}
The constructed box is equivalent (i.e. the joint probabilities are equal) to the full-correlation box defined by
\begin{equation}
P(\vec{c} \vert \vec{x}) = \begin{cases}\frac{1}{2^{n-1}}& \bigoplus\limits_{i=1}^n c_i = g_1(x_1,...,x_{k_2}) \oplus \prod\limits_{i=k_1}^{k_3}x_i \\0&\text{otherwise.}\end{cases}
\end{equation} 
\end{lemma}

\begin{IEEEproof}
The statement follows directly from the property of the full-correlation box that the set of outputs of any subset of $n-1$ parties (or smaller) is completely random \cite{BP05}, and the property that the XOR conserves randomness in case of independence.
\end{IEEEproof}

Theorem \ref{thm:comm} and Corollary \ref{cor:ungl} state that a general class of full-correlation boxes can be simulated by distillation and classical one-way communication channels. The number of these one-way channels is then smaller than the number of one-way communication channels we need if we do not apply a distillation protocol, i.e. operate from scratch. 

\begin{theorem}[Distillation with Communication]
\label{thm:comm}
Let f be a Boolean function associated to an $n$-partite full correlation box, and let f  be written as in Lemma \ref{lem:boolf}. If f fulfills $n_\mathcal{J}=1$, then:

\begin{itemize}
	\item[(i)] The full-correlation box can be constructed from generalized PR-boxes shared between a different number of parties such that in at most one generalized PR box some parties input all the time a constant.
	\item[(ii)] The number $N_{comm}^{distill}$ of necessary one-way communication channels for simulating the full-correlation box with using the generalized BS protocol is 
	\begin{equation}
	N_{comm}^{distill} \leq \begin{cases}n-1-\underset{I \in \mathcal{J}}{\max}(m_I)& \underset{I \in \mathcal{J}}{\max}(m_I) \neq n\\0& \underset{I \in \mathcal{J}}{\max}(m_I) = n. \end{cases}
	\end{equation} 
\end{itemize}
\end{theorem}

\begin{figure*}[t]
\begin{center}
\includegraphics[width=0.95\linewidth]{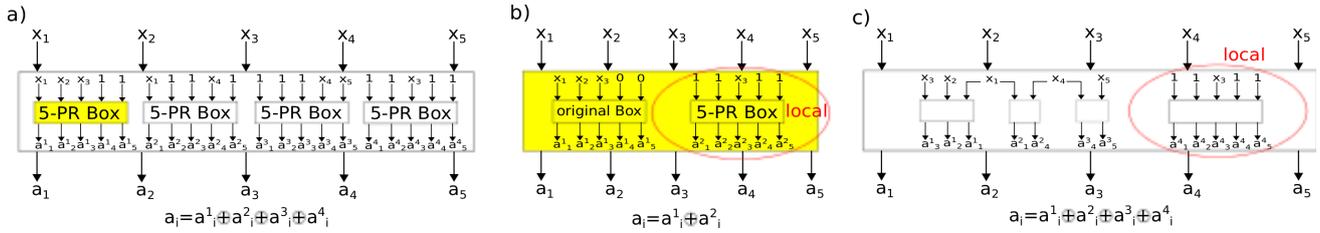} 
\end{center}
\caption{a) Simulating the full-correlation box with four 5-PR boxes. b) How to simulate the first 5-PR box with the original full-correlation box and a local box. c) Simulation of the full-correlation box with $n$-PR boxes without a constant input and a local box.}
\label{fig:example}
\end{figure*}

\begin{IEEEproof}
In this proof, we replace full-correlation boxes with $a_I = 1$ for $\vert I\vert \leq 1$ by the full-correlation box with $a_I = 0$ for $\vert I\vert \leq 1$, and all other $a_I$ for all $I \in \mathcal{I}\setminus \lbrace\emptyset\rbrace$ keep their values. We can do this by taking the XOR of the original box and the local box with $a_I = 1$ for $\vert I\vert \leq 1$. To get our original box back in the end, we take again the XOR of the changed box and the local box.

We start to prove part (i) of the theorem. The idea is to replace the boxes step by step.
In the first step, we are beginning with a $n$-PR box  with the associated set $I$. To that end, we are looking for another $n$-PR box with associated set $J$ such that $I\cap J \neq \emptyset$ (this is possible because of the assumption of the theorem). Because of Lemma~\ref{lem:constr}, we are able to replace these two boxes by two smaller boxes. We substitute the first box by an $\vert I \setminus J\vert$-PR box with inputs $I$. The second box is substituted by an $(n- \vert I\vert )$-box, where we input $J$ and for the parties $\lbrace 1, 2, ..., n\rbrace \setminus (I\cup J)$, we input 1.

Assume that we have, in this way, replaced some $n$-PR boxes by new boxes. Again, we are looking for an $n$-PR box which is not yet replaced, and whose input elements intersect with the input elements of the new box. We are making the same steps as before to replace these two boxes. In the end, we have replaced all $n$-PR boxes to a new box with the claimed properties. 

We prove part (ii) of the theorem. For this part, we assume that the replacement is made according to part (i). We have replaced the original $n$-PR boxes such that the general PR box with constant element does not correspond to the original $n$-PR box belonging to the biggest $m_I$. This is possible, since we can replace this box first. We are now able to isolate the box belonging to the biggest $m_I$. Therefore, we allow all parties that appear at least twice as well as the parties that input all the time a constant to communicate their inputs and outputs to a party which acts also in the isolated box. We have isolated the general PR box, and we are able to apply the generalized BS protocol to this box. All the other generalized PR boxes that appear in the abstraction of part (i) in the theorem can be simulated by the communication of the parties and shared randomness. So we will need $\underset{I \in \mathcal{J}}{\max}(m_I)$ one-way-communication channels less than when we start from scratch.
\end{IEEEproof}

\begin{corollary}
Let f be a Boolean function associated to an $n$-partite full correlation box, and let f  be written as in Lemma~\ref{lem:boolf}. If $n_\mathcal{J}=1$ and $\underset{I \in \mathcal{J}}{\max}(m_I) > n - \vert\bigcup\limits_{I\in \mathcal{J}} I\, \vert$ then
\begin{equation}
N_{comm}^{distill} < N_{comm}^{scratch}\ .
\end{equation}
\label{cor:ungl}
\end{corollary}

\begin{IEEEproof}
The statement follows directly from Theorems~\ref{thm:commcl} and~\ref{thm:comm}.
\end{IEEEproof}

All extremal three-partite full-correlation boxes of the non-signalling polytope fulfill the preconditions of Corollary~\ref{cor:ungl}. For more parties, it is unknown how many extremal boxes also fulfill this precondition.

\section{example}

In this example we want to distill some boxes up to the following full-correlation box: 
\begin{equation}
P(\vec{a}\vert \vec{x}) = \begin{cases} \frac{1}{2^{n-1}}&\text{$\bigoplus\limits_{i=1}^{5} a_i = x_1x_2x_3\oplus x_1x_4\oplus x_4x_5\oplus x_3$}\\0&\text{otherwise.}\end{cases}
\end{equation}

Therefore, we determine first the above-defined sets and constants. Let $\mathcal{I}= \mathcal{P}(\lbrace 1,2,3\rbrace )$. From Lemma~\ref{lem:boolf}, we know that all $a_I = 1$ for $I\in \lbrace\lbrace 1,2,3\rbrace , \lbrace 1,4\rbrace , \lbrace 4,5\rbrace , \lbrace 3\rbrace\rbrace$, and otherwise $a_I = 0$. This means that the given full-correlation box can be simulated by four 5-PR boxes with some constant inputs, where one of these boxes is local (see Fig. \ref{fig:example} a)).
We are also able to assign the set $\mathcal{J}$ of non-local $n$-PR boxes that are needed to simulate the full-correlation box:
\begin{equation}
\mathcal{J}= \lbrace\lbrace 1,2,3\rbrace , \lbrace 1,4\rbrace , \lbrace 4,5\rbrace\rbrace
\end{equation}
Each of these three non-local 5-PR boxes can be obtained from the original box by taking the XOR of the original box and the local 5-PR box when every party inputs its bits except for the parties that input the constant 1 to the 5-PR box, they input 0 in both boxes (see Fig. \ref{fig:example} b)). If we apply Theorem~\ref{thm:comm} (i), then we know that the non-local part of the original full-correlation box can be simulated by three connected $n$-PR boxes with no constant input.

Since we know $\mathcal{J}$, the number of required one-way communication channels for simulating the full-correlation box can be calculated with Theorem~\ref{thm:commcl}:
\begin{equation}
N_{comm}^{distill} =  \left|\bigcup\limits_{I\in \mathcal{J}} I \right| - 1 = 4.
\end{equation}

Obviously, this box is not local. To determine the distance (measured in the $L^1$-norm), we can use a linear program and get that the distance is 20, and the closest local box (not unique) is given by

\begin{equation}
P^L(\vec{a}\vert \vec{x}) = \begin{cases} \frac{1}{2^{n-1}}&\text{$\bigoplus\limits_{i=1}^{5} a_i = x_3$}\\0&\text{otherwise.}\end{cases}
\end{equation}

We start with the second part of the example, where we show in detail how we distill a box from the family $P_\varepsilon = \varepsilon P + (1-\varepsilon)P^L$, where $0<\varepsilon < 1$, up to $P(\vec{a}\vert \vec{x})$.

We want to distill this box arbitrarily closely to the full-correlation box above. For that, we determine first which of the parties have to communicate. Therefore, we calculate the number of parties that only belong to one of the non-local 5-PR boxes: $m_{\lbrace 1,2,3\rbrace} = 2$,
$m_{\lbrace 1,4\rbrace} =1$, and
$m_{\lbrace 4,5\rbrace} = 1$. This means that we isolate the box that belongs to the 5-PR box with three arbitrary inputs. This can be done in the same way as before: We input $(x_1, x_2, x_3, 0,0)$ in $P_\varepsilon$ and the local box and take then the XOR of its outputs. Then, we use one-way communication channels from Party 5 to 4 and one from 4 to 1. Remember that the communication channels can be used as often as the parties want. Hence, we are able to simulate perfectly the two 2-PR box, and the non-perfect 3-PR box can be isolated by communicating the inputs and outputs of the two 2-PR box to Party 1 (see Fig. \ref{fig:example} c)). We have isolated the box $P_{3,\varepsilon}^{PR}$ that is known to be distillable up to $P_3^{PR}$ by the generalized BS protocol. In this way, we are able to distill the box $P_\varepsilon$ up to the full-correlation box in the beginning.

We get that the number of one-way communication channels that is needed for this kind of distillation is $N^{distill}_{comm} = 2$, i.e., less than $N^{scratch}_{comm} = 4$.

\section{Conclusion}
We have considered the problem of non-locality distillation in the
multi-partite setting. We have found, first, that
arbitrarily weakly non-local non-isotropic approximations to
the natural
generalization of a PR box to $n$ parties are distillable by an
adaptation of a protocol for two parties. Second, this can be
applied to showing that a much more general class of extremal
correlations, including \emph{all} purely three-partite correlations, can be distilled to using {\em partial\/} communication
(less than if no weak systems can be used). In this context, weak
non-locality, hence, manages to replace communication between a
subset of parties. It remains a challenging open problem to
understand,
classify, and apply multi-party non-locality better. It seems that for
certain tasks (such as randomness amplification), multi-party
non-locality
outperforms bipartite correlations.

\section*{Acknowledgment}
The authors thank Jibran Rashid, Benno Salwey, Marcel Pfaffhauser and
Daniela Frauchiger for helpful discussions.
This work was supported by the Swiss National Science Foundation (SNF),
the NCCR "Quantum Science and Technology" (QSIT), and the COST
action on "Fundamental Problems in Quantum Physics."


\end{document}